\documentclass{jltp} 
\usepackage{graphicx,amssymb}

\title{Soliton-limited Superflow in $^{3}$He-A between Parallel 
Plates}
\author{J. Kopu and E.~V.~Thuneberg}
\address{Low Temperature Laboratory, Helsinki University of 
Technology, Finland}
\runninghead{J. Kopu and E.~V.~Thuneberg}{Soliton-limited superflow in
$^{3}$He-A between parallel  plates}
\begin{document}

\maketitle

\begin{abstract}
We have studied theoretically the flow of superfluid $^{3}$He-A in
parallel-plate geometry. The
equilibrium order-parameter texture is calculated numerically in
two spatial dimensions consisting of the coordinates along the flow
direction and perpendicular to the plates. The calculations have been
done using the hydrostatic theory in the Ginzburg-Landau region and
assuming a large external magnetic field perpendicular to the plane of
calculation. We have studied a uniform texture and a dipole-unlocked
splay soliton as initial configurations. In the former case we find
the Freedericksz transition and a helical instability with increasing
flow. In the latter case we find instability in the soliton. This
instability is closely related to the critical velocity in the
presence of a vortex sheet. Also, the transverse NMR frequency shift
at the soliton has been calculated.

\end{abstract}
\def\tensor#1{\stackrel{\leftrightarrow}{#1}}

\section{Introduction}

The anisotropic A phase of superfluid $^{3}$He ex\-hibits unique
behavior under externally applied superflow. Instead of being
quantized, the circulation of the superfluid velocity ${\bf v}_{\rm
s}$ in $^{3}$He-A can have any value depending on the spatial
variation of the order-parameter texture. The decay of superflow can
be accomplished by generating a nonuniform texture of the orbital
vector $\hat{\bf l}$, leading ultimately to the formation of
continuous vortices (with no singular core). However, energy
considerations favor the uniform texture with constant $\hat{\bf l}$
for small velocities and postpone these processes until a certain
critical velocity is exceeded. Because of the macroscopic
length scale of texture variations and the boundary condition that
fixes the direction of $\hat{\bf l}$ at the walls, vortex formation is
unaffected by extrinsic influences, such as surface roughness and
thermal and quantum fluctuations. Therefore, in contrast to
superfluid $^{4}$He, critical velocity in $^{3}$He-A is determined by
an intrinsic instability of the flow.

Superflow of $^{3}$He-A has been previously studied theoretically
in many papers, see Refs.~\onlinecite{Voll} and \onlinecite{1d} for
references. Essentially all of them consider only
one-dimensional flow. Here we present the first truly two-dimensional
calculation for $^{3}$He-A in the \mbox{presence} of external flow.
We study a flow channel consisting of parallel
plates.  With increasing flow, an
initially constant texture  becomes first modified in the Freedericksz
transition, where
$\hat{\bf l}$ starts to tilt from the direction normal to the
plates\cite{DeGe,Fetter}.  On increasing the velocity further, the
texture becomes unstable towards a helical
deformation\cite{Bhatta,Lin-Liu}, which leads to formation of vortices.
However, the main point in this paper is to study an initially
inhomogeneous texture formed by a dipole-unlocked soliton\cite{Kumar}.
The flow through such a domain wall was first studied by Vollhardt and
Maki\cite{Maki} using a variational ansatz. More recently the same
problem was studied numerically in Ref.\ 
\onlinecite{1d}. We extend these one-dimensional calculations to
include the effect of lateral walls in a flow channel. We investigate
how the critical velocity in the presence of the soliton depends on
the width $D$ of the flow channel. We find that the dependence is much
stronger than for the helical instability in the absence of the
soliton.   

The dipole-locked soliton is exceptionally interesting, because it
forms the backbone of another topological object, the vortex
sheet\cite{Erkki}. Our calculation can be used as a model for the
critical velocity of the sheet, and it explains semi-quantitatively
the measured dependence on the angular velocity\cite{VSExp}.  Peculiar
dynamical properties of the vortex sheet was observed in
experiments\cite{Eltsov}. To provide additional information to explain
this behavior, we have also calculated numerically the NMR frequency
shift which is the relevant experimental parameter. The shifts are
determined as functions of flow velocity and the separation between the
plates.

\section{Hydrostatic theory}
\label{theory}

Our starting point is very similar to that in Ref.~\onlinecite{1d}. 
Here we repeat the main points. The A-phase order parameter
$\tensor{A}$ is a 3$\times$3 tensor defined by two orthogonal unit
vectors $\hat{\bf m}$ and $\hat{\bf n}$ in the orbital space and a unit
vector $\hat{\bf d}$ in the spin space and has the form\cite{Voll} 
(with real $\Delta$)
\begin{equation}
A_{\mu j} = \Delta \hat{d}_{\mu} (\hat{m}_{j} + i\hat{n}_{j}).
\end{equation}
One defines ${\hat{\bf l}}\equiv {\hat{\bf m}}\times{\hat{\bf n}}$
and a superfluid velocity as
\begin{equation}
{\bf v}_{\rm s}= 
{\hbar\over 2m_3}\sum_j\hat
m_j \nabla\hat n_j, 
\label{supvel}
\end{equation}
where $m_3$ is the mass of a $^{3}$He atom. 
The equilibrium form of the order-parameter texture corresponds to a
minimum of the hydrostatic free energy. In order to avoid dealing with
complicated constraints between $\hat{\bf l}$ and ${\bf v}_{\rm s}$,
we express the free energy in terms of the unit-vector fields
$\hat{\bf m}({\bf r})$, $\hat{\bf n}({\bf r})$, and $\hat{\bf d}({\bf
r})$. For simplicity, we restrict all our calculations to the 
Ginzburg-Landau region ($T_{\rm c}-T \ll T_{\rm c}$). 
The energy density consists of 
the dipole-dipole energy
\begin{equation}
f_{\rm d} = \textstyle{1\over 2}
\lambda_{\rm d}\Big[(\hat{\bf d}\cdot\hat{\bf m})^2+
(\hat{\bf d}\cdot\hat{\bf n})^2\Big],
\label{fdip}
\end{equation}
the magnetic-field energy
\begin{equation}
f_{\rm h} = \textstyle{1\over 2}
\lambda_{\rm h}(\hat{\bf d}\cdot{\bf H})^2,
\label{fhoo}
\end{equation}
and the gradient energy
\begin{eqnarray}
&f_{\rm g}&= \frac{\hbar^2\rho_{\parallel}}{16m_3^2}
\bigg\{ \sum_{ik}\Big[(\nabla_{i}\hat m_k)^{2}+
(\nabla_{i}\hat n_k)^{2}\Big] +(\gamma-1)
\Big[(\nabla\cdot\hat{\bf m})^2
\nonumber \\
&&\mbox{}+(\nabla\cdot\hat{\bf n})^2
+\sum_i (\hat{\bf m}\cdot\nabla\hat
d_i)^{2} + \sum_i(\hat{\bf n}\cdot\nabla\hat d_i)^{2} \Big]
+2\sum_{ik}(\nabla_{i}\hat d_k)^{2} \bigg\}, 
\label{fg}
\end{eqnarray}
see Ref.~\onlinecite{Janne} for details. Generally, the superfluid
density is anisotropic with components $\rho_\parallel$ and $\rho_\perp$
corresponding to the flow parallel and perpendicular to $\hat{\bf l}$,
respectively. In the Ginzburg-Landau region with the weak-coupling
value $\gamma=3$, they differ by a factor of two:
$\rho_\perp=2\rho_\parallel$. 
The supercurrent density is given by
\begin{eqnarray}
j_{s,k}&=&\frac{\hbar\rho_{\parallel}}{4m_3}
 \sum_j\Big[
(\gamma-2)(\hat m_k\nabla_{j}\hat n_j-\hat n_k\nabla_{j}\hat m_j) 
\nonumber \\
&&\mbox{}+\hat m_j\nabla_{k}\hat n_j-\hat n_j\nabla_{k}\hat m_j 
+\hat m_j\nabla_{j}\hat n_k-\hat n_j\nabla_{j}\hat m_k\Big]. 
\end{eqnarray}
A natural unit for distance is the dipole
length $\xi_{\rm d}=(\hbar/2m_3)\sqrt{\rho_\parallel /\lambda_{\rm d}}$. 
Similarly, we define \mbox{units} for velocity $v_{\rm d} \equiv
\sqrt{\lambda_{\rm d}/\rho_\parallel}$
and 
current $j_{\rm d} \equiv \rho_\parallel v_{\rm d}$. 
The free energy is minimized under the
con\-straint that $\hat{\bf l}$ must be parallel to the surface normal
$\hat{\bf s}$ at the walls. 
We restrict ourself in all following calculations to the large-field
limit ($H \gg H_{\rm d} \equiv
\sqrt{\lambda_{\rm d}/\lambda_{\rm h}}$), where $\hat{\bf d}$ is locked
to the plane perpendicular to the direction of ${\bf H}$.

\begin{figure}[tb]
\begin{center}\leavevmode
\includegraphics[width=0.6\linewidth]{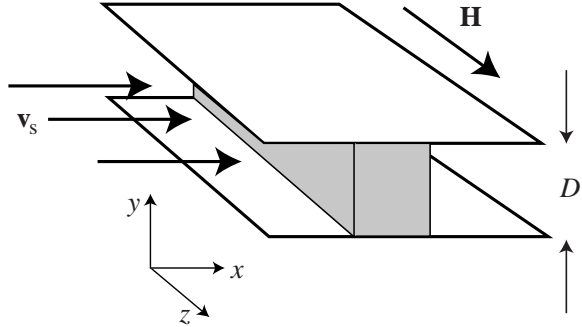}
\caption[geometry]{ 
The parallel-plate geometry used in the calculations. The separation
between the plates
is $D$ and the superfluid velocity ${\bf v}_{\rm s}$
is in the positive $x$ direction. 
The orientations of the external magnetic field ${\bf H}$
and the soliton wall (grey) are also shown.
}\label{geometry}\end{center}\end{figure}
A schematic view of the situation we wish to describe is presented in
Fig.~\ref{geometry}. In our calculations we consider a rectangular
area in the $xy$ plane with a length $L$ in the direction of the flow
($-L/2<x<L/2$) and width $D$ in the $y$ direction ($-D/2<y<D/2$). We
assume translational invariance in the $z$ direction. The boundary
condition at the plate surfaces has the form $\hat{m}_{y}=
\hat{n}_{y}=0$. Due to the presence of a large magnetic
field ${\bf H}\parallel\hat{\bf z}$, $\hat{d}_z=0$.

In the numerical calculation
the order parameter is defined at $N\times M$
discrete points spaced by $\Delta x$ and $\Delta y$ in $x$ and $y$
directions, respectively ($N\Delta x=L$, $M\Delta y=D$). 
The discretized free energy functional of Eqs.~(\ref{fdip})-(\ref{fg}) 
transforms into a function of the values of $\hat{\bf
m}$, $\hat{\bf n}$ and $\hat{\bf d}$ at the lattice points.
The minimum of this function is then sought by the following method:
first an initial guess for the order-parameter vectors is chosen.
Then the texture is changed towards the equilibrium distribution
according to
\begin{eqnarray}
\Delta\hat{\bf d}&=&
\epsilon_{\rm d} 
\mbox{\boldmath $\tau$}_{\rm d} 
\times \hat{\bf d},
\nonumber \\ 
\Delta\hat{\bf m}&=&
\epsilon_{\rm o} 
\mbox{\boldmath $\tau$}_{\rm o} 
\times \hat{\bf m},
\nonumber \\ 
\Delta\hat{\bf n}&=&
\epsilon_{\rm o} 
\mbox{\boldmath $\tau$}_{\rm o} 
\times \hat{\bf n},
\end{eqnarray}
where the torques acting on the order-parameter vectors
are defined as
\begin{eqnarray}
\mbox{\boldmath $\tau$}_{\rm d} &\equiv&
\frac{\delta F}{\delta \hat{\bf d}} \times \hat{\bf d},
\nonumber \\ 
\mbox{\boldmath $\tau$}_{\rm o} &\equiv&
\frac{\delta F}{\delta \hat{\bf m}} \times \hat{\bf m}+
\frac{\delta F}{\delta \hat{\bf n}} \times \hat{\bf n},
\end{eqnarray}
and the (small) 
iteration constants $\epsilon_{\rm d}$ and $\epsilon_{\rm o}$
are chosen so as to achieve fast convergence. 
Also, after each iteration step the vectors are adjusted to have
unit length and to satisfy the condition $\hat{\bf m} \perp \hat{\bf n}$.
The iteration process is repeated
until the minimum of the free energy is reached (signalled by
the vanishing of the torques).

We generate the superflow in our model
by imposing a fixed difference $\Phi(x=L/2)-\Phi(x=-L/2)$ 
$\equiv \Delta\Phi$, which will be defined more precisely below. 
For differ\-ent values of $\Delta\Phi$,
we monitor the average induced supercurrent density
\begin{equation}
j \equiv \langle j_{s,x} \rangle=\frac{1}{D}\int_{-D/2}^{D/2} {\rm d}y 
\hspace{1mm} j_{s,x}(x,y),
\end{equation}
which is independent of $x$ in the converged solution.
We study the current-velocity relationship $j(v)$,
where $v \equiv (\hbar/2m_3 L)\Delta\Phi$ 
can be interpreted as the velocity of the normal component
corresponding to the phase difference $\Delta\Phi$ (Ref.\
\onlinecite{1d}).  The critical velocity $v_{\rm c}$ of the flow
instability is determined from the condition
\begin{equation}
\frac{\partial j}{\partial v} \vert_{v=v_{\rm c}} = 0.
\end{equation}

\section{Uniform initial state}
\label{uniform} 

Before studying the soliton case, it is necessary to understand
the flow response in the absence of the soliton.
We begin by considering the simplest order-parameter structure
in $^{3}$He-A between parallel plates.  
For the case of zero velocity, $v=0$, the free energy is minimized
by a uniform texture, where $\hat{\bf l}\parallel\hat{\bf d}\parallel\hat{\bf y}$.
This homogeneous configuration is a simultaneous minimum of $f_{\rm g}$,
$f_{\rm d}$, and $f_{\rm h}$, and also satisfies the requirement 
$\hat{\bf l}\parallel\hat{\bf s}$ at the boundaries. 
Next we introduce a flow in the system by a requirement
\begin{eqnarray}
\tensor{A}(x=L/2,y)&=&e^{i\Delta\Phi}\tensor{A}(x=-L/2,y) \nonumber \\
&=&e^{2ivm_3 L/\hbar}\tensor{A}(x=-L/2,y).
\label{bc1}
\end{eqnarray}
In the presence of flow, $\hat{\bf l}$ tends to turn towards the flow
direction because the component of the superfluid-density tensor along
$\hat{\bf l}$, $\rho_\parallel$, is smaller than the perpendicular
one, $\rho_\perp$. This would require the formation of an
inhomogeneous texture, because at the surfaces of the plates $\hat{\bf
l}$ is rigidly anchored perpendicular to the flow.  Therefore, for
small enough velocities the uni\-form texture
$\hat{\bf l}\perp{\bf v}_{\rm s}$ persists as the equilibrium
configuration, with $\hat{\bf m}$ and $\hat{\bf n}$ turning around
constant $\hat{\bf l}$ to achieve the required phase difference or,
equivalently, $\tensor{A}=$ const.$\times
\exp(2ivm_3 x/\hbar)$. 

The transition where $\hat{\bf l}$
starts to deflect from the plate normal $\hat{\bf y}$ is known as 
Freedericksz transition, and $v_{\rm Fr}$ denotes the corresponding
velocity. The Freedericksz transition has been studied extensively in
the case where it is induced by magnetic field ${\bf H}\parallel\hat{\bf
y}$\cite{Hook}. In the present case, it was first calculated by
deGennes and Rainer\cite{DeGe}, who neglected the dipole-dipole
interaction (\ref{fdip}) and obtained
\begin{equation}
v_{\rm Fr} = \frac{c}{D},
\label{fre}
\end{equation}
where $c=(\pi\hbar/2m_3)\sqrt{3/4}$. 
This result holds in the limit $D\ll\xi_{\rm d}$, but in a wider slab
there is additional textural rigidity because $\hat{\bf d}$ tends
to follow $\hat{\bf l}$. This has been calculated both exactly and by
variational ansatz by Fetter\cite{Fetter}. In the limit $D\gg\xi_{\rm
d}$ there is complete dipole-locking and one obtains (\ref{fre})
with
$c=(\pi\hbar/2m_3)\sqrt{5/4}$. At intermediate $D$'s $v_{\rm Fr}$
monotonically interpolates between these limits (see Fig.~\ref{jd}
below).

\begin{figure}[tb]
\begin{center}\leavevmode
\includegraphics[width=0.7\linewidth]{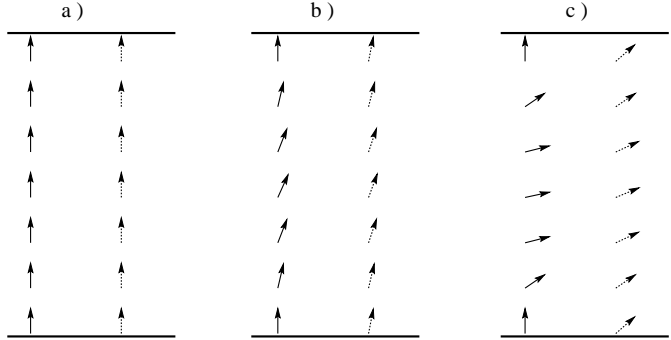}
\caption[uni]{ 
The variation of $\hat{\bf l}$ (solid arrows) and $\hat{\bf d}$ (dashed arrows) 
for $D=6\xi_{\rm d}$, when $v_{\rm Fr}\approx0.51 v_{\rm d}$. Both
fields are translationally invariant in $x$ direction. The textures are
for a) $v<v_{\rm Fr}$ (homogeneous texture), b) $v=0.55 v_{\rm d}$,
and c) $v=1.3 v_{\rm d}$.
}\label{uni}\end{center}\end{figure}
\begin{figure}[tb]
\begin{center}\leavevmode
\includegraphics[width=0.6\linewidth]{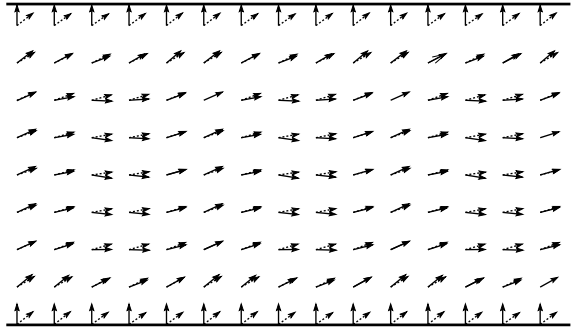}
\caption[helical]{ 
Helical disturbances of $\hat{\bf l}$ (solid arrows) and $\hat{\bf d}$
(dashed arrows)
for $D=8\xi_{\rm d}$ at $v=1.3 v_{\rm d}$. 
A length of $L=15\xi_{\rm d}$ in the direction
of flow is presented, comprising three wavelengths of the helix.
}\label{helical}\end{center}\end{figure}

Typical $\hat{\bf l}$ and $\hat{\bf d}$ textures for different values of $v$
are presented in Fig.~\ref{uni}. 
On increasing the velocity further beyond $v_{\rm Fr}$,
$\hat{\bf l}$ bends more and more until
most of the texture (excluding the vicinity of the surfaces)
has aligned itself with the flow direction. Finally, 
at $v=v_{\rm h}$,
the texture
becomes unstable towards helical disturbances that break
the translational invariance in the $x$ coordinate\cite{Bhatta},
see Fig.~\ref{helical}.
We have not made systematic studies of this transition here, but it
seems that there is no substantial difference to the corresponding
one-dimensional calculation\cite{1d}, which corresponds to the limit
$D\rightarrow\infty$. The critical velocity increases slightly because
of the presence of the lateral walls,
\begin{equation}
v_{\rm h}(D) \approx v_{\rm h}(\infty)+\frac{a}{D^2},
\label{hel}
\end{equation}
where $v_{\rm h}(\infty)=1.26v_{\rm d}$ (Ref.\ \onlinecite{1d}) and
$a\approx 3v_{\rm d}\xi_{\rm d}^2$. Another question is the
stability of these helical textures. It has been found in one
dimension that the helical textures are unstable towards nucleating
vortices in the vicinity of $T_{\rm c}$\cite{Lin-Liu,1d}. Whether this is true also
in our restricted geometry remains open. 

\begin{figure}[tb]
\begin{center}\leavevmode
\includegraphics[width=0.45\linewidth]{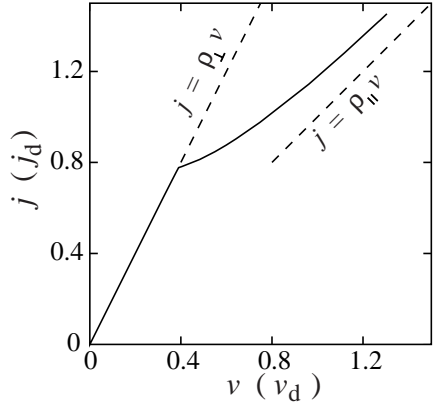}
\caption[curvel]{ 
The current-velocity diagram for $D=8\xi_{\rm d}$. The Freedericksz
transition shows up as a kink at $v\approx 0.4v_{\rm d}$ and the curve
ends to the onset of helical texture at $v\approx 1.3v_{\rm d}$. The
lines
$j=\rho_\perp v$ and
$j=\rho_\parallel v$ are shown as guides to the eye.
}\label{curvel}\end{center}\end{figure}

A useful way of expressing the response an externally applied flow
is through the current-velocity diagram $j(v)$. A typical example of 
such a diagram is presented in Fig.~\ref{curvel}. At low velocities
the uniform texture with $\hat{\bf l}\perp{\bf v}_{\rm s}$ can be
seen as a linear slope $j=\rho_\perp v$. 
At the Freedericksz transition part of the $\hat{\bf l}$ texture
starts to deflect towards ${\bf v}_{\rm s}$, causing a 
decrease in the slope of $j$. With increasing velocity the
texture tends to align with ${\bf v}_{\rm s}$ and the slope approaches
$\rho_\parallel$. 

\section{Flow with a soliton}
\label{soliton} 

\begin{figure}[tb]
\begin{center}\leavevmode
\includegraphics[width=0.6\linewidth]{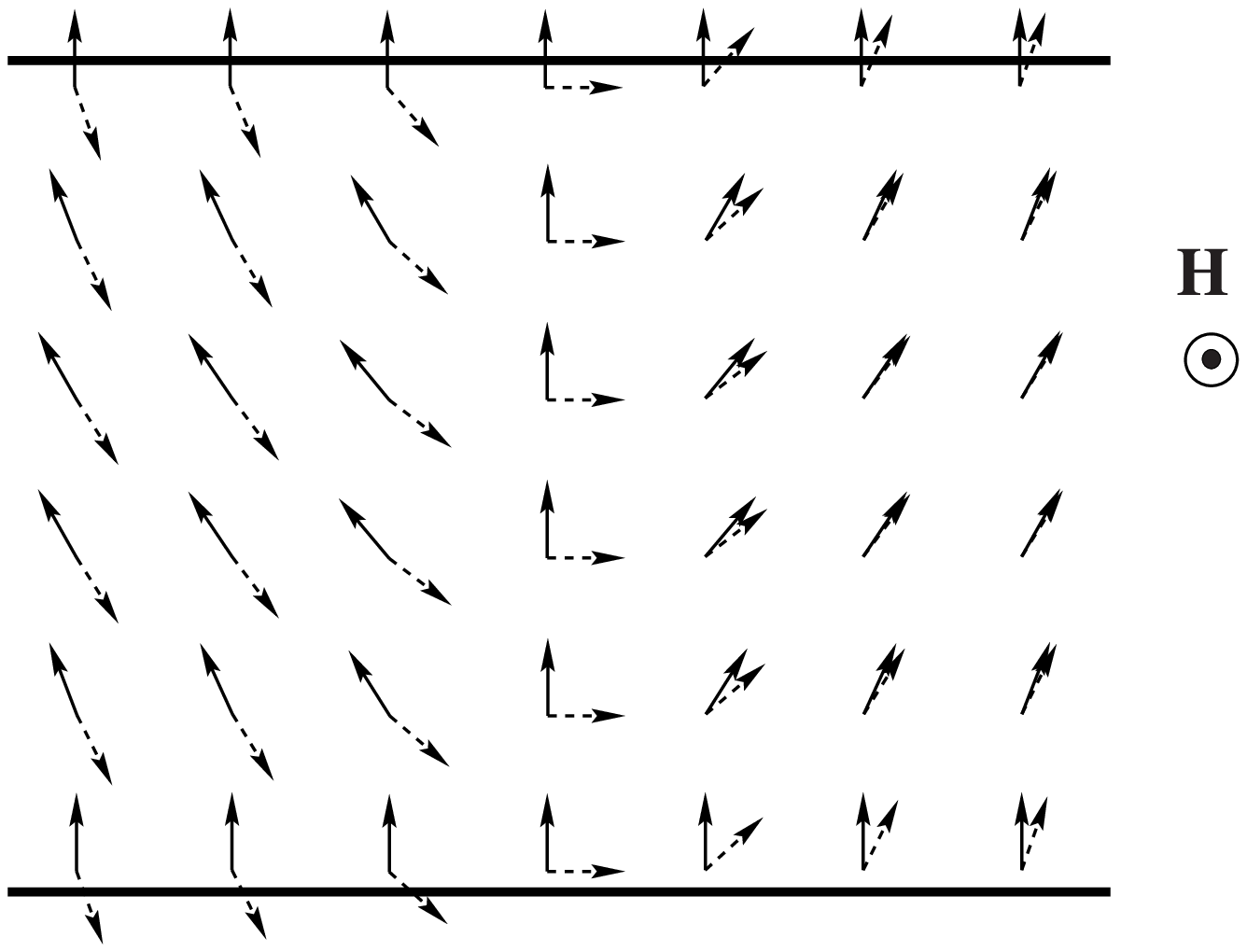}
\caption[ldvect]{ 
The variation of $\hat{\bf l}$ (solid arrows)
and $\hat{\bf d}$ (dashed arrows)
in a splay soliton structure for $D=8\xi_{\rm d}$ and zero velocity. 
On the left side $\hat{\bf l}=-\hat{\bf d}$
and on the right side $\hat{\bf l}=\hat{\bf d}$.
}\label{ldvect}\end{center}\end{figure}

The complicated order-parameter structure of $^{3}$He-A allows for a
variety of different topological defects. One variety of these are
solitons, i.e. two-dimensional domain-wall-like structures separating
two energetically degenerate minima of the dipole-dipole interaction
(\ref{fdip}). On one side of the soliton $\hat{\bf l}$ and $\hat{\bf
d}$ are parallel to each other, on the other side antiparallel. 
In our geometry the soliton has the so-called splay structure, i.e.
the magnetic field is in the plane of the soliton,
see Fig.~\ref{ldvect}. Because of the orienting effect of
the walls, far away from the soliton a uniform $\hat{\bf l}=\hat{\bf
y}$ texture is approached, with $\hat{\bf d}=-\hat{\bf y}$ on the left
side and $\hat{\bf d}=\hat{\bf y}$ on the right side of the wall.

The boundary condition (\ref{bc1}) we used to generate flow 
in the uniform case
is not consistent with the presence of a soliton. Instead, it is
possible to show that the combined operation of translation by $L$ in
the flow direction and reflection in the $y$ coordinate 
is a symmetry operation for the soliton texture. Therefore, a flow
velocity equal to $v$ can be achieved by requiring
\begin{equation}
\tensor{A}(x=L/2,y)=e^{2ivm_3 L/\hbar}
\sigma_y \tensor{A}(x=-L/2,-y)\sigma_y,
\label{bc2}
\end{equation}
where $\sigma_y$ is a diagonal matrix with elements 1, -1, and 1.
Otherwise the numerical calculations follow the same procedure as
in the case with a uniform initial texture.

\begin{figure}[tb]
\begin{center}\leavevmode
\includegraphics[width=0.7\linewidth]{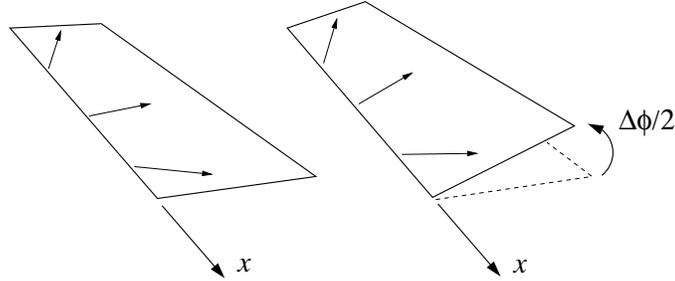}
\caption[sol]{ 
Flow response of a one-dimensional soliton structure (schematic). 
A phase difference of
$\Delta\Phi$ can be accomplished by rotating texture an angle of
$\Delta\Phi/2$ around the flow direction.
}\label{sol}\end{center}\end{figure}

In order to understand the response of the soliton wall to an external
flow, it is advantageous to first study a simpler case in one
dimension $x$ along the flow in the bulk. We consider the range from
$x=-L/2$ to $x=L/2$, over which the direction of $\hat{\bf l}$ is reversed
from $\hat{\bf l}=-\hat{\bf x}$ to $\hat{\bf l}=\hat{\bf x}$, see
Fig.~\ref{sol}. In the absence of a magnetic field, it is possible to
show that, starting with a texture with zero phase difference over
the length
$L$, a texture with a phase difference equal to $\Delta\Phi$ can be
formed by simply rotating the initial texture an angle $\Delta\Phi/2$
around $\hat{\bf x}$. Furthermore, also the final texture carries zero
supercurrent. The situation changes if a magnetic field ${\bf H}\perp
\hat{\bf x}$ is introduced, because it prevents $\hat{\bf l}$ from
rotating freely through the dipole-dipole interaction. As a result, a
texture corresponding to a nonzero supercurrent is formed. When the
phase difference exceeds a critical value, the texture undergoes an
abrupt rotation through a large angle, again leading to vortex
creation somewhere on the sides. 
Roughly the same scenario takes place also in the two-dimensional
parallel-plate geometry, with some modifications. Far away from the
soliton plane and at the solid walls the texture is locked to a uniform
$\hat{\bf l}=\hat{\bf y}$ configuration and cannot rotate when the
flow is applied. This results in an additional rigidity opposing
the rotation of the soliton. 

\begin{figure}[tb]
\begin{center}\leavevmode
\includegraphics[width=0.5\linewidth]{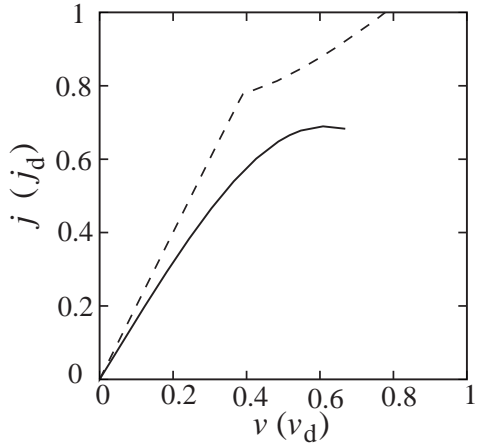}
\caption[solcur]{ 
The current-velocity diagram for a splay soliton texture for
$D=L=8\xi_{\rm d}$ (solid line). The diagram in the absence
of the soliton
for a channel of the same width is also shown (dashed line). 
}\label{solcur}\end{center}\end{figure}

\begin{figure}[tb]
\begin{center}\leavevmode
\includegraphics[width=0.6\linewidth]{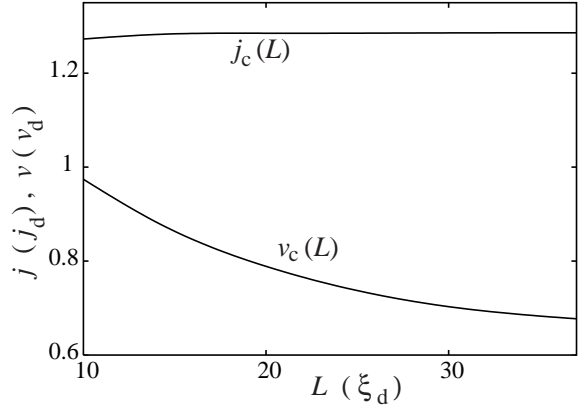}
\caption[converge]{ 
The critical velocity $u_{\rm c}$ and current $j_{\rm c}$
for a \mbox{splay} soliton
as functions of $L$ for $D=4\xi_{\rm d}$.  
}\label{converge}\end{center}\end{figure}

The calculation of the critical velocity for the soliton
presents some problems because, due to the breakdown of
translational invariance in the $x$ direction, the flow
response of the system depends on the length $L$ of our computational 
region. We define the critical velocity as
\begin{equation}
v_{\rm c} \equiv \lim_{L\rightarrow \infty} 
(\hbar/2m_3)\frac{\Delta\Phi_{\rm c}(L)}{L},
\end{equation}
where $\Delta\Phi_{\rm c}$ corresponds to the phase difference for
which the maximum current $j_{\rm c}(L)$ is achieved. A typical form
of the current-velocity diagram 
at finite $L$ is shown in Fig.~\ref{solcur}. With
increasing $L$ (thus increasing the relative length of the
soliton-free region) the current-velocity curve approaches the
translationally invariant behavior discussed in the previous chapter.
However, it turns out that $v_{\rm c}(L)$ shows appreciable dependence
on
$L$ even for $L \gg \xi_{\rm d}$. On the other hand, the value of the
critical current $j_{\rm c}(L)$ only varies on a much smaller scale (on
the order of a few percent). The dependences are presented in
Fig.~\ref{converge}. Therefore, in order to avoid lengthy calculations
involving a huge number of lattice points, we calculate the critical
current using a moderate-sized lattice and read the corresponding
velocity $v_{\rm c}$ from the current-velocity curve without the
soliton (such as in Fig.~\ref{curvel}). 

\begin{figure}[tp]
\begin{center}\leavevmode
\includegraphics[width=0.6\linewidth]{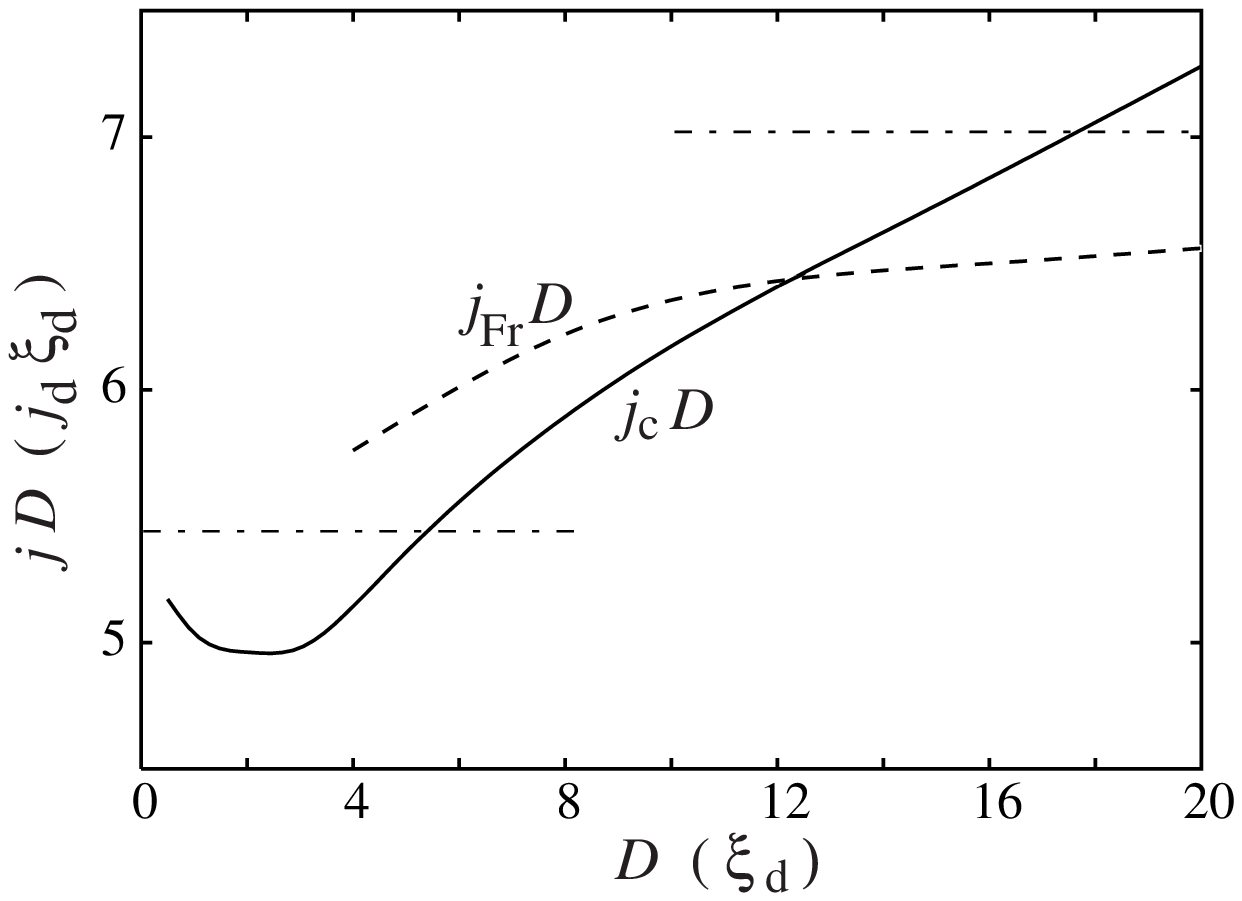}
\caption[jd]{ 
The critical current $j_{\rm c}$ times the plate separation $D$
for a splay soliton as a function of $D$ (solid line)
Also shown is $j_{\rm Fr}D$ corresponding to 
the Freedericksz transition of a uniform texture (dashed line), together with 
analytic results (\ref{fre})
for the limiting cases of dipole-locking (upper dash-dotted line)
and for $\hat{\bf d}$ = constant (lower dash-dotted line).
}\label{jd}\end{center}\end{figure}

\begin{figure}[tp]
\begin{center}\leavevmode
\includegraphics[width=0.6\linewidth]{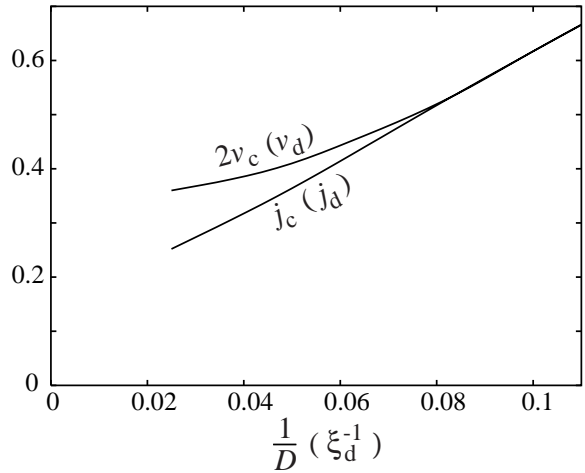}
\caption[kriittiset]{ 
The critical current $j_{\rm c}$ and twice the critical velocity
$2v_{\rm c}$
for a splay soliton as a function of $1/D$.
}\label{kriittiset}\end{center}\end{figure}

The critical current $j_{\rm c}$ for the splay soliton texture is
presented in Figs.~\ref{jd} and~\ref{kriittiset} as a function of the
plate separation $D$. In Fig.~\ref{jd} we have plotted the product
$j_{\rm c}D$. 
We have also included for comparison the numerically calculated
$j_{\rm Fr}D$ and the analytic limits 
(\ref{fre})
for the Freedericksz transition. In order to have a more accurate look at
the limit of large plate separations, we present $j_{\rm c}$ and
$2v_{\rm c}$ as functions of inverse $D$ in Fig.~\ref{kriittiset}.
As long as the critical currents exceed $j_{\rm Fr}
\equiv 2v_{\rm Fr}$ (i.e. when $D \lesssim 12 \xi_{\rm d}$), 
the dimensionless
critical currents and velocities are simply related through
a factor of two:
$j_{\rm c}/j_{\rm d}=2v_{\rm c}/v_{\rm d}$.
However, for large plate separations $j$
exceeds $j_{\rm Fr}$, and 
the critical velocity decreases more slowly
(see Fig.~\ref{curvel}). 

From the figures we find that, for
intermediate values of $D$, $j_{\rm c}$ obeys reasonably well the
relation
\begin{equation}
j_{\rm c} \approx \frac{a}{D}+b,
\end{equation}
with $a=4.5j_{\rm d}\xi_{\rm d}$, $b=0.17j_{\rm d}$ for $D \lesssim 12
\xi_{\rm d}$, and
$a=5.1j_{\rm d}\xi_{\rm d}$, $b=0.11j_{\rm d}$ for $D > 12\xi_{\rm d}$.
Fig.~\ref{kriittiset} suggests that $v_{\rm c}$ levels off at the limit
of large $D$ to $0.15v_{\rm d}<v_{\rm c}^{\rm bulk}
<0.2v_{\rm d}$,
in agreement with our 
previous result obtained from a one-dimensional treatment 
of a dipole-unlocked soliton\cite{1d}. Note also that 
$v_{\rm c}/v_{\rm d}=j_{\rm c}/j_{\rm d}$ in the limit
$D \rightarrow \infty$. The main conclusion from the calculations
above is that the critical velocity in the presence of a soliton
is small and depends essentially on the width $D$ of the
channel. The dependence is much stronger than for the helical
instability in the absence of the soliton.

The calculation above gives a qualitative explanation for the measured
critical velocity of a vortex sheet. The vortex sheet consists of a
soliton backbone to which continuous vorticity is added\cite{Erkki}. 
In a rotating container the sheet has end lines at the side walls of the
container. During angular acceleration there is counterflow through the
vortex-free end pieces of the sheet, and the sheet can grow only when
new vorticity is nucleated there. It is found experimentally that the
critical velocity for the growth of the vortex sheet depends
essentially on the angular velocity $\Omega$, increasing with
increasing $\Omega$\cite{VSExp}. This can now
be understood because the width of the vortex-free soliton pieces
decreases with increasing $\Omega$, which corresponds to decreasing
$D$ in a channel and thus increasing $v_{\rm c}$.  The channel model is
still a rather crude approximation for the end piece of a vortex sheet,
but in spite of that it gives the right order of magnitude for the
critical current and its dependence on $\Omega$.

We can argue analogously for the continuous vortex lines. Since we found
only weak dependence of the helical instability on the channel width
(\ref{hel}), the critical velocity for creating vortex lines should
be essentially independent on the rotation velocity. This is what is
found experimentally\cite{VSExp}.

\section{Nuclear magnetic resonance}
\label{nuclear}

NMR techniques are effective in obtaining information about
the order-parameter distribution in superfluid $^{3}$He.
The frequency of the transverse NMR absorption peak is given by\cite{NMR,Voll,Erkki}
\begin{equation}
\omega^2=\omega_0^2+R_\perp^2\omega_\parallel^2, 
\end{equation}
where $\omega_0$ is
the Larmor frequency, $\omega_\parallel$ the A-phase longitudinal
resonance frequency, and $R_\perp^2$ is determined by solving
the eigenvalue problem\cite{NMR}
\begin{equation}
-(2\nabla^2-\nabla\cdot\hat{\bf l}\hat{\bf l}\cdot\nabla)\psi+
U\psi=(R_\perp^2-1)\psi
\end{equation} 
for the transverse fluctuations of $\hat{\bf d}$ given by the
wave function $\psi$. 
The potential for the fluctuations is given by
\begin{equation}
U=-\hat{l}_z^2-(\hat{\bf l}\times\hat{\bf d})^2-2(\nabla\theta)^2
+(\hat{\bf l}\cdot\nabla\theta)^2,
\label{pot}
\end{equation}
where $\hat{\bf d}=\hat{\bf x}\cos\theta+\hat{\bf y}\sin\theta$
corresponds to the unperturbed distribution in the presence of
a large magnetic field ${\bf H}\parallel\hat{\bf z}$. 
An approximation for the lowest eigenvalue and the corresponding 
bound state of fluctuation induced by the rf field 
can be obtained by the following variational principle:
\begin{equation}
R_\perp^2-1= \min_\psi
\frac{\int {\rm d}^3{\bf r} (2\vert\nabla\psi\vert^2-
\vert \hat{\bf l}\cdot\nabla\psi\vert^2+U\vert\psi\vert^2)}
{\int {\rm d}^3{\bf r}\vert\psi\vert^2}.
\label{r2}
\end{equation}
For a uniform state with 
$\hat{\bf d}\parallel\hat{\bf l}\perp{\bf H}$,
the potential in Eq. (\ref{pot}) vanishes, $\psi$ is constant and
the absorption occurs at the bulk frequency given by $R_{\perp}^2=1$.
However, dipole-unlocked regions such as the splay soliton define
a potential well, and usually give rise to the presence of bound
states with frequencies corresponding to $R_{\perp}^2<1$. 
The effective volume of absorption associated with a given mode
in a homogeneous rf field is given by
\begin{equation}
V_{\rm NMR}=\frac{\vert\int {\rm d}^3{\bf r} \psi({\bf r})
\exp[-i\theta({\bf r})]\vert^2}{\int {\rm d}^3{\bf r}\vert\psi\vert^2}.
\label{vnmr}
\end{equation}
In the lowest mode $\vert \psi \vert^{2}$ is expected to accumulate \mbox{near} 
the minima of the potential energy $U$, which in our problem occur
at the two points where the soliton joins the plates.
Due to the broken reflection symmetry in $y \rightarrow -y$
($\hat{\bf s}\cdot\hat{\bf l}$ changes sign at the wall,
see Fig.~\ref{ldvect}) two effects arise: the potential wells
have different depths, and the directional derivatives 
$\vert\hat{\bf l}\cdot\nabla\psi\vert^2$
of Eq.~(\ref{r2}) give different contributions near the two joining
points. As a result, one of the points is singled
out as $D$ is increased.

\begin{figure}[tp]
\begin{center}\leavevmode
\includegraphics[width=0.6\linewidth]{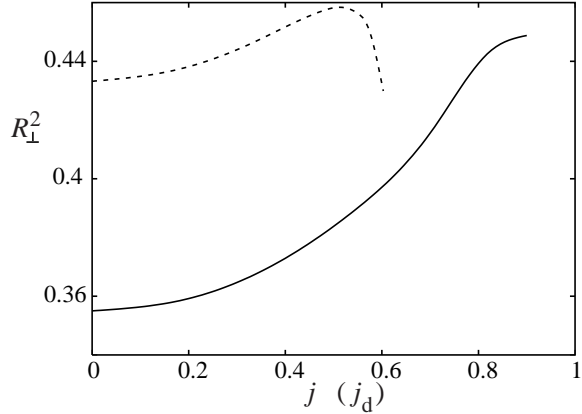}
\caption[shift]{ 
Transverse NMR frequency shift $R_\perp^2$ as a function of
the current $j$ for two different plate separations
of 6$\xi_{\rm d}$ (solid line) and 10$\xi_{\rm d}$ (dashed line).
}\label{shift}\end{center}\end{figure}

Recent experiments on vortex sheet dynamics have exhibited unexpected
behavior\cite{Eltsov}. A possible explanation involves a rotating
\mbox{state} that consists of several sheets and multiple contact
points with the container. The sheets are joined to the walls by
vorticity-free soliton pieces. An important question arising in the
explanation is whether these pieces have the same resonance frequencies
than the sheet itself. Figure \ref{shift} shows $R_{\perp}^2$ as a
function of the flow velocity for two different plate separations. The
frequencies we obtained are considerably lower than those associated
with bulk splay soliton, calculated e.g. in Ref.~\onlinecite{Kumar} for
zero velocity as $R_\perp^2=0.677$. In fact, the frequencies are rather
close to the ones expected for a vortex sheet, previously calculated to
be
$R_\perp^2=0.46$--0.48, depending on the model\cite{Erkki}.

\begin{figure}[tb]
\begin{center}\leavevmode
\includegraphics[width=0.6\linewidth]{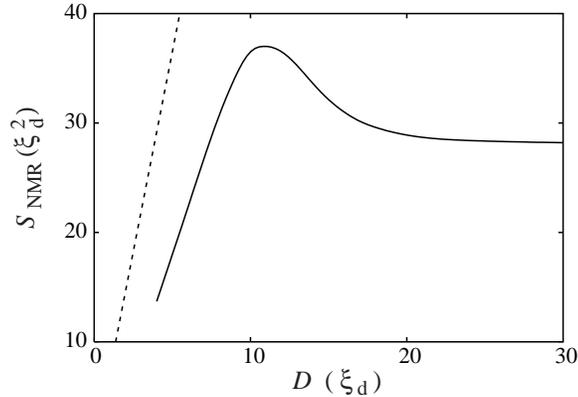}
\caption[snmr]{ 
The effective area of absorption $S_{\rm NMR}$ of the lowest
eigenmode
as a function of the plate separation $D$ for zero velocity (solid line).
Also shown is the area of absorption for a bulk splay soliton
of length $D$ (dashed line).
}\label{snmr}\end{center}\end{figure}

In addition to the frequency shift associated to a given mode, another
important quantity is the relative intensity of the mode. This can be
expressed in terms of the effective volume of absorption, $V_{\rm
NMR}$, as defined in \mbox{Eq.~(\ref{vnmr})}.  For small values of
$D$, $\vert\psi\vert^2$ is distributed almost evenly along the entire
soliton, and $V_{\rm NMR}$ grows linearly with increasing $D$. After
$D$ is increased enough, $\vert\psi\vert^2$ concentrates at the
joining point of the soliton and one of the plates with $y=-D/2$
($\hat{\bf s}\cdot\hat{\bf l}=1$), resulting in a constant $V_{\rm
NMR}$.  Figure \ref{snmr} shows the effective area of absorption
(volume per unit length in $z$ direction) for zero phase difference,
$v=0$. For comparison, we have also drawn the area of absorption for a
bulk splay soliton of length $D$, calculated to have a absorption
thickness of 7.3$\xi_{\rm d}$\cite{Risto}. The absorption thickness
of a vortex sheet is approximately $6$--$8 \xi_{\rm d}$ depending on the
angular velocity\cite{Erkki}.

\section{Conclusions}
\label{conc}

We have studied the flow stability of superfluid $^{3}$He-A in the
vicinity of solid walls by calculating numerically the order-parameter
distribution in two spatial dimensions. The emphasis of this work was
to study the flow in the presence of a splay composite soliton. We
have determined the critical current and velocity for the creation of
continuous vortices as functions of the separation between the
walls. The results give a qualitative explanation to the
experimentally measurable critical velocity for the growth of a vortex
sheet. The transverse NMR frequency shift $R_{\perp}^2$ for the lowest
eigenmode of the dipole-unlocked soliton was found to be rather close
to that of a vortex sheet.


\end{document}